\documentclass[10pt, conference, letterpaper]{IEEEtran}
\IEEEoverridecommandlockouts
\usepackage{cite}
\usepackage{amsmath,amssymb,amsfonts}
\usepackage{algorithmic}
\usepackage{algorithm}
\usepackage{graphicx}
\usepackage{textcomp}
\usepackage{xcolor}
\usepackage{booktabs} 

\graphicspath{{picture/}}

\def\BibTeX{{\rm B\kern-.05em{\sc i\kern-.025em b}\kern-.08em
    T\kern-.1667em\lower.7ex\hbox{E}\kern-.125emX}}

\begin{document}

\title{Efficient and Accurate Surrogate-Assisted Electromagnetic Parameter Calibration for 6G Digital Twin Channels}

\author{\IEEEauthorblockN{Xiaofan Zou, Pan Tang, Peijie Liu, Changyou Tai}
\IEEEauthorblockA{Beijing University of Posts and Telecommunications, Beijing, China\\
E-mail: \{zouxf, tangpan27\}@bupt.edu.cn}
}

\maketitle

\begin{abstract}
High-fidelity ray-tracing (RT) channel reconstruction is a fundamental step toward building digital twins for the era of 6G wireless communications. However, precise calibration of complex electromagnetic material parameters remains a dual challenge characterized by massive computational overhead and strict accuracy requirements. To overcome this bottleneck, we propose a Surrogate-assisted Grey Wolf Optimizer with Individual Memory (SGWO-IM) algorithm that simultaneously improves computational efficiency and calibration accuracy. In terms of computational efficiency, an online surrogate model is seamlessly embedded into the evaluation workflow for candidate pre-screening, substantially reducing the reliance on highly time-consuming real RT simulations. Regarding calibration accuracy, adaptive convergence and individual memory strategies are incorporated to optimize the global parameter search path, effectively enhancing the consistency between the reconstructed channel and measured data. Validated against measured channel data from a high-density urban scenario, the proposed algorithm requires only 225 real RT simulation calls compared to the 600 calls needed by the standard Grey Wolf Optimizer (GWO), cutting computational overhead by 62.5\%. Concurrently, the final Root Mean Square Error (RMSE) is substantially reduced from the 3.65 dB of GWO to 2.97 dB. The results demonstrate that the SGWO-IM algorithm achieves significant advancements in both efficiency and precision, providing a solution that effectively balances efficiency and accuracy for electromagnetic environment reconstruction.
\end{abstract}

\begin{IEEEkeywords}
Digital Twin Channel, Ray Tracing, Electromagnetic Parameter Calibration, Surrogate Model, Grey Wolf Optimizer, V2V Communications
\end{IEEEkeywords}

\section{Introduction}
With the evolution of the sixth-generation (6G) mobile communication technology, the paradigm of wireless communication is shifting towards proactive environment intelligence communication \cite{b1}. In this context, building large-scale channel foundation models, such as ChannelGPT \cite{b2}, requires massive and highly accurate digital twin channel data. Therefore, ray tracing (RT) has become a core method for the high-fidelity reconstruction of complex electromagnetic (EM) environments due to its capability to precisely depict EM wave propagation paths \cite{b3,b4,b5,b6,b7,b8}. However, the accuracy of RT models largely depends on the accurate input of EM parameters (such as permittivity and conductivity) of various building materials within the real scenario. Since these parameters vary significantly with frequency bands and are difficult to measure directly, utilizing measured channel data for reverse calibration has become crucial \cite{b9,b10}. Essentially, this is a highly challenging multi-variable, non-convex, and computationally expensive non-linear optimization task.

To address these multi-variable calibration challenges, various optimization algorithms have been explored \cite{b11}. Traditional heuristic and swarm intelligence algorithms possess global search capabilities \cite{b12}, but they encounter a severe computational bottleneck in practical deployments \cite{b13}. These methods require executing time-consuming forward RT simulations for the entire population during every iteration, resulting in massive computational overhead. Moreover, due to the highly coupled nature of channel parameters, standard algorithms often suffer from a loss of population diversity, causing the optimization to prematurely stop at suboptimal points \cite{b14}. Recently, machine learning methods have been introduced to accelerate the process by replacing simulation engines with neural networks. However, constructing a reliable static surrogate model offline still requires enormous initial computational resources to generate massive training datasets \cite{b15,b16,b17}. Consequently, a calibration framework that can simultaneously achieve high efficiency and high precision without extensive offline data preparation remains highly desired.

To overcome these limitations, this paper proposes a Surrogate-assisted Grey Wolf Optimizer with Individual Memory (SGWO-IM) for highly efficient and accurate digital twin channel reconstruction. The main contributions of this paper are summarized as follows:
\begin{itemize}
    \item A surrogate-assisted calibration framework is proposed to eliminate computational bottlenecks: To avoid the massive overhead of standard swarm intelligence algorithms, an online Radial Basis Function (RBF) neural network is introduced as a dynamic surrogate model. By utilizing initial samples for low-cost pre-screening, expensive forward RT simulations are strictly restricted to the predicted elite individuals. Validated in a real-world scenario, this approach significantly accelerates the optimization process, reducing real RT simulation calls by 62.5\%.
  \item A dynamic search mechanism is designed to enhance parameter estimation precision: The proposed algorithm incorporates a non-linear cosine convergence factor alongside an individual cognitive component. This structural refinement overcomes the standard Grey Wolf Optimizer (GWO)'s limitation of blindly following the current best solution, ensuring a thorough exploration of the parameter space. Validated by empirical Vehicle-to-Vehicle (V2V) measurement data from a typical urban street scenario, this mechanism substantially improves the parameter estimation accuracy, reducing the final Root Mean Square Error (RMSE) from 3.65 dB of the standard GWO to 2.97 dB, thereby enabling the RT model to accurately fit the actual path loss.
\end{itemize}

The remainder of this paper is organized as follows. Section II details the proposed SGWO-IM algorithm. Section III introduces the measurement and simulation settings. Section IV presents the performance evaluation and results analysis. Finally, Section V concludes this paper.

\section{Proposed Method}

\subsection{Overall Process}
\begin{figure*}[htbp]
\centerline{\includegraphics[width=0.8\textwidth]{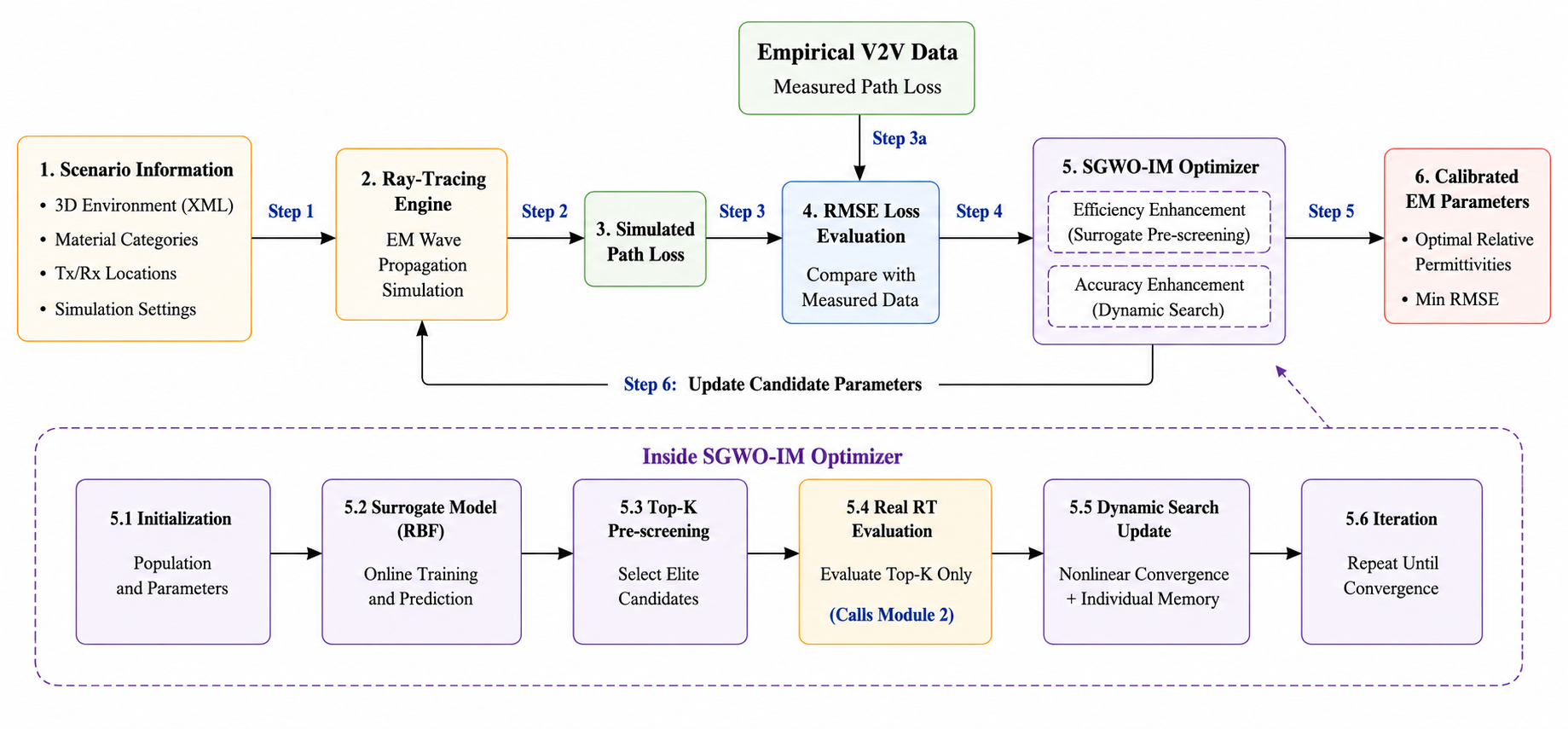}}
\caption{Flowchart of the proposed SGWO-IM parameter calibration framework.}
\label{fig1}
\end{figure*}

As shown in Fig. \ref{fig1}, the overall execution process of the proposed parameter calibration approach is structured as a continuous feedback loop. Initially (Step 1), the comprehensive scenario information, including the XML-based 3D environment, material categories, Tx/Rx locations, and specific simulation settings, is fed into the RT engine. Subsequently (Steps 2 and 3), the RT engine executes the EM wave propagation simulation to output the simulated path loss (PL), while the empirical V2V data is concurrently imported to serve as the real-world measured PL baseline (Step 3a). 

To quantify the discrepancy between the simulated results and the measured data, the Root Mean Square Error (RMSE) evaluation (Step 4) is conducted using the following objective function:
\begin{equation}
\mathrm{RMSE} = \sqrt{\frac{1}{M} \sum_{m=1}^{M} (L_{m}^{\mathrm{RT}} - L_{m}^{\mathrm{Meas}})^2},
\end{equation}
where $M$ is the total number of measurement points, and $L_{m}^{\mathrm{RT}}$ and $L_{m}^{\mathrm{Meas}}$ represent the simulated PL by the RT engine and the actual measured PL at the $m$-th receiver location, respectively. 

Driven by this error feedback, the SGWO-IM optimizer iteratively updates the candidate parameters (Step 6). Internally, it utilizes an RBF surrogate model for pre-screening to enhance efficiency, and applies a dynamic search mechanism to improve parameter estimation accuracy. The optimizer then generates an updated set of candidate EM parameters, which are fed back into the RT engine for the next evaluation. This closed-loop process repeats until the convergence criteria are satisfied, ultimately outputting the calibrated EM parameters—specifically, the optimal relative permittivities—that yield the minimum RMSE (Step 5).

\subsection{Proposed SGWO-IM Algorithm for Parameter Calibration}
When calibrating EM parameters, traditional swarm intelligence algorithms face two major challenges: the severe computational bottleneck caused by massive forward RT simulations, and the risk of premature convergence due to the highly coupled nature of channel parameters. To address these issues, the proposed SGWO-IM algorithm optimizes the standard framework across two primary dimensions: computational efficiency and calibration accuracy. The detailed procedure is summarized in Algorithm \ref{alg1}.

\subsubsection{Computational Efficiency Enhancement via Online Surrogate}
In traditional optimization, every candidate parameter combination in the population (size $N$) must be evaluated by the RT engine during each iteration. To eliminate this massive computational overhead, an RBF neural network is integrated into the framework as an online surrogate model to predict the evaluation errors.

In the early iterations, all candidate parameters are evaluated using the real RT simulator to accumulate reliable initial data samples. Once the iteration threshold is reached, these historical samples are utilized to train the RBF network online. In the subsequent iterations, the RBF model first predicts the RMSE for all $N$ candidate combinations at a negligible computational cost. Based on these predictions, the candidates are sorted, and only the top-$K$ elite individuals (set to $K=5$ in this paper) with the lowest predicted errors are selected for the actual RT simulation. This pre-screening mechanism significantly reduces the real RT simulation calls per iteration from $N$ to $K$, fundamentally alleviating the computational burden without compromising the final calibration accuracy.

\subsubsection{Calibration Accuracy Enhancement via Dynamic Search}
To prevent the algorithm from stagnating at suboptimal parameter combinations, a dynamic search mechanism is designed to enhance the accuracy of the parameter calibration. This mechanism consists of a non-linear cosine convergence factor and an individual historical memory component.

First, standard algorithms use a linearly decreasing convergence factor, which struggles to adapt to the complex bounds of the EM parameter space. Therefore, a non-linear factor $a(t)$ is introduced:
\begin{equation}
a(t) = 2 \cdot \cos \left( \frac{\pi}{2} \cdot \frac{t}{T_{max}} \right),
\end{equation}
where $t$ is the current iteration and $T_{max}$ is the maximum number of iterations. This strategy maintains a large search step size in the early stage, ensuring a thorough exploration across the entire parameter space (e.g., the wide permittivity bounds of concrete and glass). In the later stage, it decays rapidly to facilitate fine-tuning around the optimal values.

Second, to address the standard algorithm's limitation of merely following the current best solution, an individual historical best ($\mathrm{Pbest}$) memory mechanism is incorporated \cite{b18}. During the parameter update process, each candidate solution retains the memory of the best parameter combination it has historically discovered. The position update formula is thereby modified as:
\begin{equation}
\vec{X}_{new}(t+1) = (1 - w(t)) \cdot \vec{X}_{\mathrm{GWO}}(t+1) + w(t) \cdot \vec{X}_{\mathrm{Pbest}}(t),
\end{equation}
where $\vec{X}_{\mathrm{GWO}}(t+1)$ is the standard guidance position, and the dynamic memory weight $w(t)$ is defined as:
\begin{equation}
w(t) = w_{min} + (w_{max} - w_{min}) \cdot \left( \frac{t}{T_{max}} \right)^2.
\end{equation}
In this study, $w_{min}$ and $w_{max}$ are set to 0.1 and 0.3, respectively. Through this dynamic fusion, the candidate parameters not only move towards the global optimal region but also refer to their independent historical optimums. This continuous adjustment ensures the spatial diversity of the candidate solutions, directly improving the algorithm's capability to find the precise EM parameters.
\begin{algorithm}[htbp]
\caption{SGWO-IM Algorithm}
\label{alg1}
\begin{algorithmic}[1]
\STATE \textbf{Step 1: Initialization}
\STATE Initialize the population, maximum iterations, and bounds.
\STATE Initialize individual memory $\vec{X}_{\mathrm{Pbest}}$ and $w$.
\STATE Initialize positions and scores of $\alpha$, $\beta$, and $\delta$ wolves.
\STATE Initialize an empty training set for the RBF surrogate model.
\STATE \textbf{Step 2: Iterative Optimization}
\FOR{$t = 1$ \TO $T_{max}$}
    \IF{RBF model is available}
        \STATE Predict RMSE for all $N$ candidates using the RBF model.
        \STATE Sort and select top-$K$ elites to form evaluation subset.
    \ELSE
        \STATE Select all $N$ candidates (Evaluate entire population).
    \ENDIF
    \FOR{\textbf{each} candidate in subset}
        \STATE Call RT engine to calculate actual RMSE.
        \STATE Add $(X, \mathrm{RMSE})$ to the RBF training set.
        \IF{$\mathrm{RMSE} < \mathrm{Pbest\_score}$}
            \STATE Update personal best: $\vec{X}_{\mathrm{Pbest}} = X$.
        \ENDIF
        \STATE Update the positions of $\alpha$, $\beta$, and $\delta$ wolves.
    \ENDFOR
    \STATE Online retrain and update the RBF surrogate model.
    \STATE Calculate non-linear convergence factor $a(t)$.
    \STATE Calculate dynamic memory weight $w(t)$.
    \FOR{$i = 1$ \TO $N$}
        \STATE Calculate standard GWO position $\vec{X}_{\mathrm{GWO}}$ guided by $\alpha, \beta, \delta$ and factor $a$.
        \STATE Update position via memory fusion using $\vec{X}_{\mathrm{GWO}}$ and $\vec{X}_{\mathrm{Pbest}}$.
    \ENDFOR
\ENDFOR
\STATE \textbf{Step 3: Output}
\STATE Return the optimal EM parameters and minimum RMSE.
\end{algorithmic}
\end{algorithm}

\section{Measurement and Simulation Settings}
\begin{figure}[htbp]
\centerline{\includegraphics[width=0.8\columnwidth]{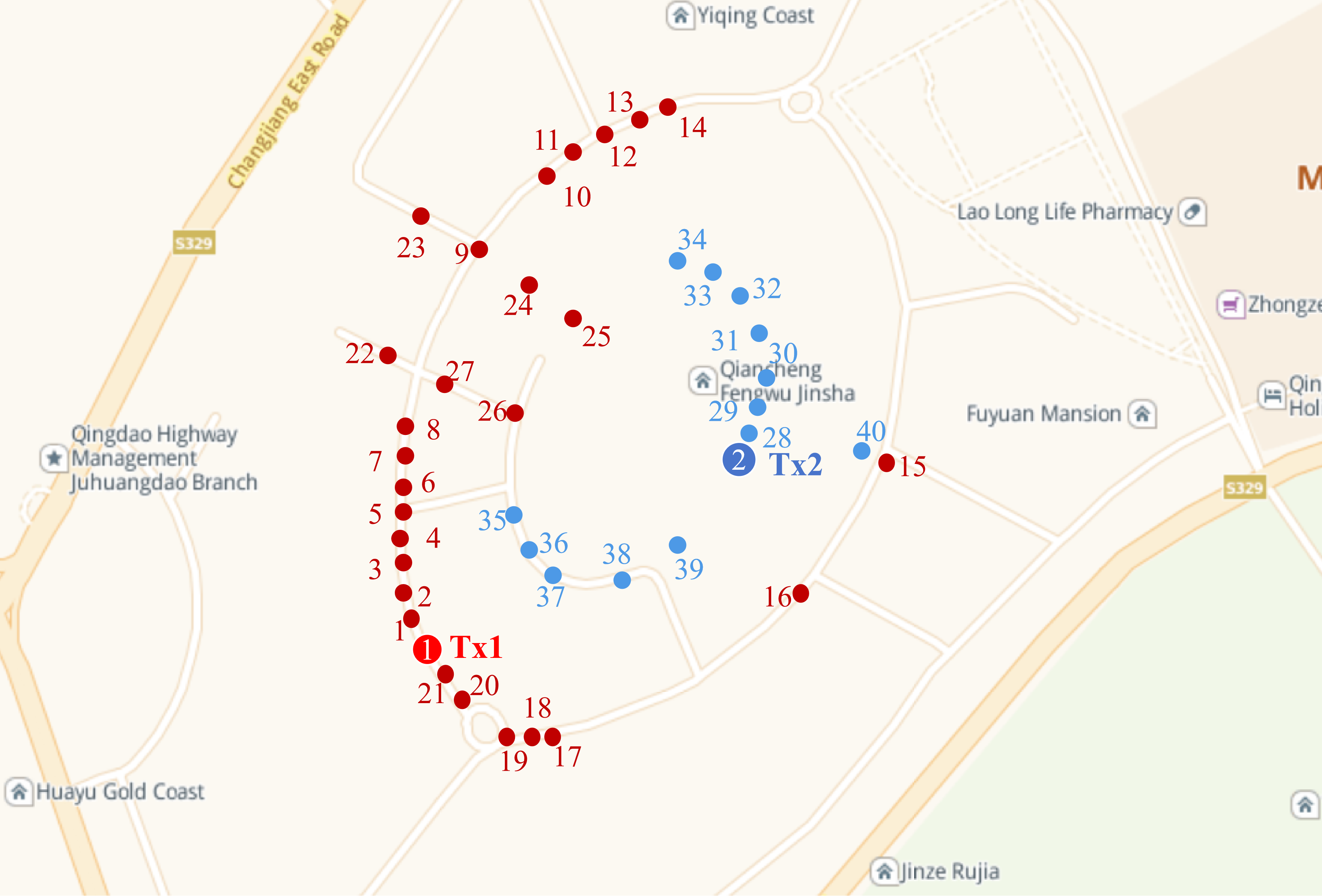}}
\caption{Schematic diagram of measurement point locations.}
\label{fig2}
\end{figure}

\begin{figure}[htbp]
\centerline{\includegraphics[width=0.8\columnwidth]{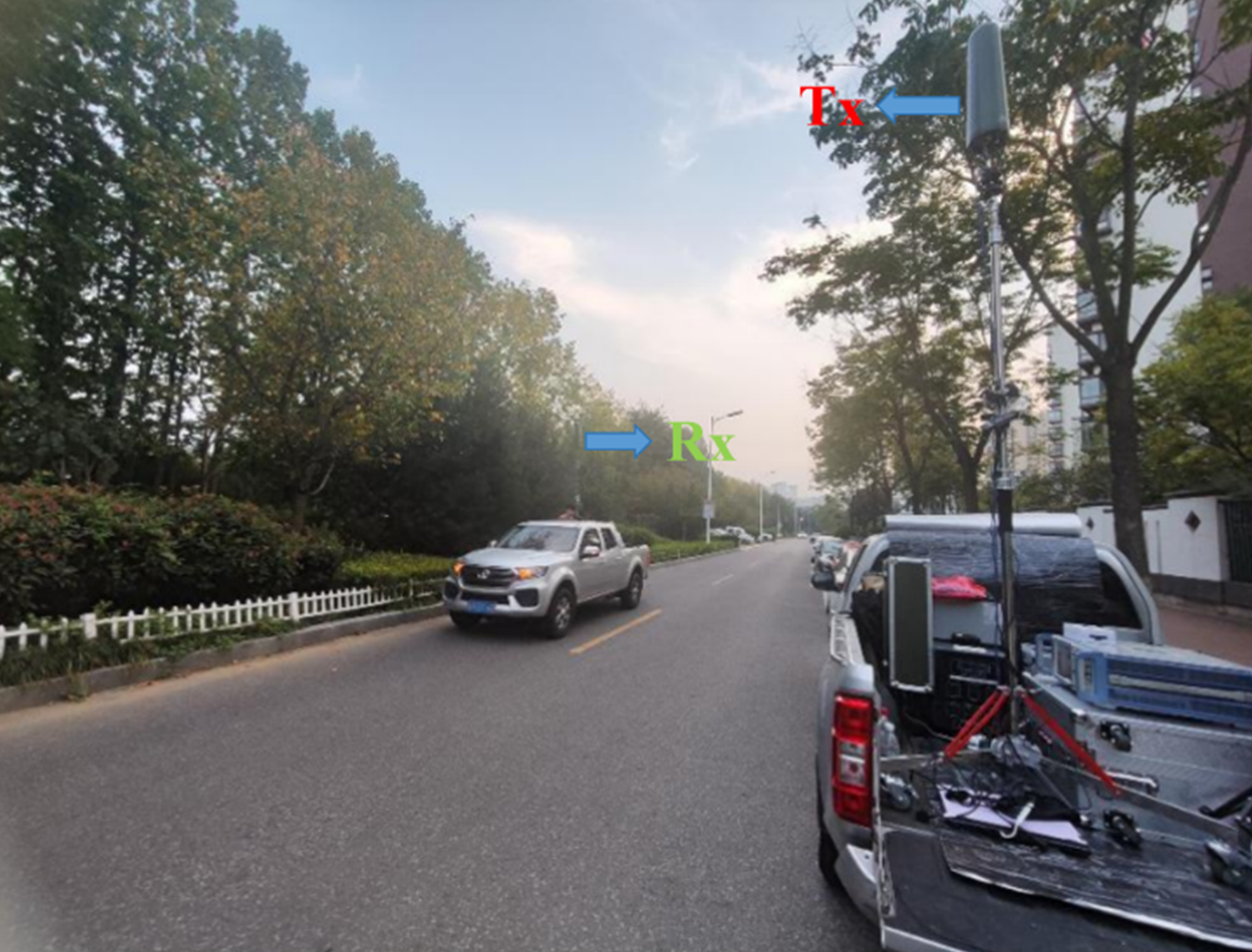}}
\caption{V2V channel measurement setup and hardware deployment.}
\label{fig3}
\end{figure}

The channel measurement experiment in this study was conducted in a high-density urban street scenario near Golden Beach, Huangdao District, Qingdao, covering a geographical area of approximately 2.1 km $\times$ 1.7 km. This measurement campaign follows the established methodology for urban V2V channel characterization \cite{b19}, ensuring reliability in capturing complex multipath propagation features. The buildings in this area generally have heights ranging from 60 to 70 m, featuring concrete main structures with numerous glass windows on the facades. The underlying surfaces are primarily composed of soil and asphalt, while discrete scatterers such as street lamp poles and green belts are randomly distributed along the streets. During the measurement, the Tx was mounted in the rear bed of a pickup truck and served as a fixed static node, while the Rx vehicle moved along the streets and stopped at specific locations for fixed-point data collection. The experiment involved two fixed Tx locations, Tx 1 and Tx 2. The Tx was initially fixed at the Tx 1 position while the Rx vehicle conducted fixed-point measurements at 27 distinct receiver locations labeled as Rx 1 to 27 distributed in the surrounding area. After completing these designated point measurements, the Tx was relocated and fixed at the Tx 2 position, and the Rx vehicle proceeded to measure the remaining 13 locations labeled as Rx 28 to 40 to continue data acquisition, as shown in Fig. \ref{fig2}. Throughout the process, a Global Positioning System (GPS) module was utilized to synchronously record the instantaneous coordinates and timestamps of both transceivers, ensuring precise alignment between spatial distances and channel data. To ensure measurement consistency and acquire sufficient samples, the Rx continuously recorded 100 sets of channel impulse response data at each fixed measurement point. The heights of both omnidirectional antennas were strictly maintained at 4 m above the ground. The specific measurement scenario and hardware deployment are illustrated in Fig. \ref{fig3}.

The transmission system consists of a vector signal generator (R\&S SMW 200A), a power amplifier, and an omnidirectional antenna. The signal generator transmits a periodic 511-sample pseudo-noise sequence 9 (PN9) with a transmit power of -10 dBm. The receiving end is equipped with an omnidirectional antenna, a low-noise amplifier, a spectrum analyzer (R\&S FSW 43), and a laptop for data processing. Relevant measurement parameters are listed in Table \ref{tab1}.

\begin{table}[htbp]
\caption{Measurement Parameters}
\begin{center}
\begin{tabular}{cc}
\toprule
\textbf{Parameters} & \textbf{Value/Type} \\
\midrule
Carrier frequency & 4 GHz \\
RF bandwidth & 200 MHz \\
Tx/Rx Antenna Type & Omnidirectional / Omnidirectional \\
Tx Antenna Height & 4 m \\
Rx Antenna Height & 4 m \\
Tx Signal & PN9 \\
\bottomrule
\end{tabular}
\label{tab1}
\end{center}
\end{table}

To achieve high-fidelity simulation of realistic radio wave propagation, a forward RT simulation platform was developed based on the MATLAB programming environment. The platform first imports the geometric data file in .xml format of the study area, along with the coordinates of the 40 valid spatial sample points obtained from the actual fixed-point measurements, comprising 27 receiver points corresponding to Tx 1 and 13 receiver points corresponding to Tx 2. Subsequently, the RT platform systematically searches for all EM wave propagation paths between the Tx and Rx based on geometric optics. After identifying these paths, the system utilizes their geometric features to perform EM calculations, determining the field strength distribution along each path. Finally, it extracts the channel characteristics to complete the entire simulation process \cite{b20}. Following this, by calculating the RMSE between the simulated PL and the measured PL for each point, the proposed SGWO-IM algorithm is employed to optimize and calibrate the EM parameters of key materials. To validate the performance of the proposed SGWO-IM algorithm, the standard GWO was selected as the baseline algorithm. Detailed settings are shown in Table \ref{tab2}.

\begin{table}[htbp]
\caption{Algorithm Parameter Settings}
\begin{center}
\begin{tabular}{ccc}
\toprule
\textbf{Parameters} & \textbf{SGWO-IM} & \textbf{Standard GWO} \\
\midrule
Max iterations & 20 & 20 \\
Population size (N) & 30 & 30 \\
Convergence strategy & Non-linear cosine & Linear decay \\
Memory weight & Quadratic (0.1--0.3) & N/A \\
Eval. iterations & 5 & N/A \\
RT evaluations/iter & 5 (Top K) & 30 (Full population) \\
\bottomrule
\end{tabular}
\label{tab2}
\end{center}
\end{table}

\section{Results and Analysis}

\subsection{Computational Efficiency Analysis}
\begin{figure}[htbp]
\centerline{\includegraphics[width=\columnwidth]{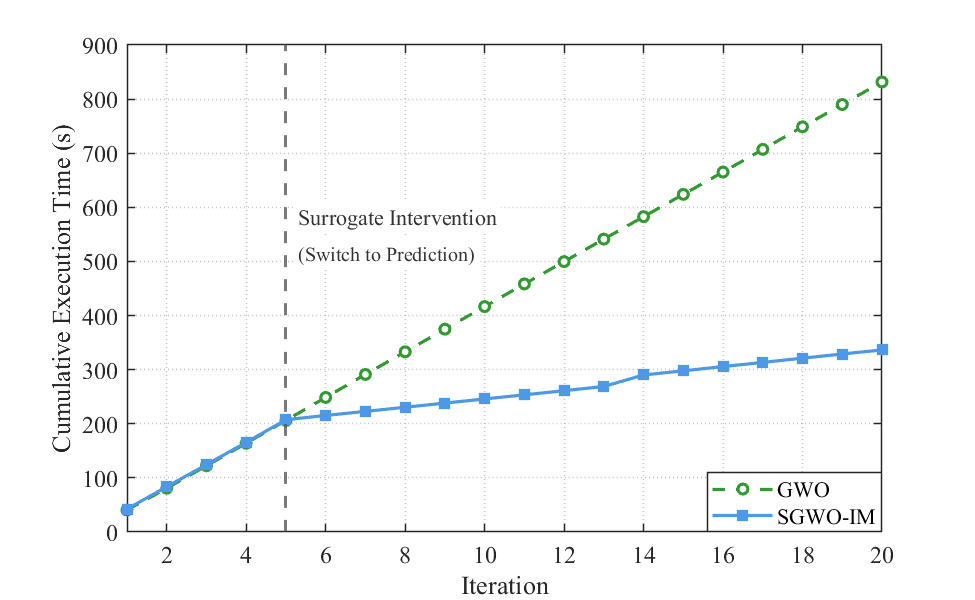}} 
\caption{Comparison of cumulative execution time between GWO and SGWO-IM.}
\label{fig_eff}
\end{figure}

Computational overhead represents a major bottleneck in RT-based channel calibration. To evaluate the acceleration capability of the proposed algorithm, Fig. \ref{fig_eff} illustrates the cumulative execution time over 20 iterations, and Table \ref{tab_eff} provides a quantitative comparison of the computational costs.

\begin{table}[htbp]
\caption{Comparison of Computational Efficiency}
\begin{center}
\begin{tabular}{lcc}
\toprule
\textbf{Algorithm} & \textbf{RT Simulation Calls} & \textbf{Final Cumulative Time (s)} \\
\midrule
GWO & 600 & 831.68 \\
SGWO-IM & 225 & 336.51 \\
\bottomrule
\end{tabular}
\label{tab_eff}
\end{center}
\end{table}

As shown in Fig. \ref{fig_eff} and Table \ref{tab_eff}, the GWO exhibits a steep linear growth in computational cost, requiring 600 time-consuming RT simulations to complete the process, which results in a total execution time of 831.68 s. In contrast, the cumulative cost curve of SGWO-IM flattens significantly after the 5th iteration. The vertical dashed line in Fig. \ref{fig_eff}, explicitly labeled as ``Surrogate Intervention", marks the critical transition point where the RBF surrogate model is sufficiently trained and takes over the evaluation process. By restricting the computationally expensive RT evaluations only to the top 5 elites during the remaining 15 iterations, SGWO-IM reduces the total RT simulations to 225. Consequently, the final cumulative time drops to 336.51 s. This demonstrates that SGWO-IM cuts the overall computational overhead by 62.5\% in terms of RT calls and saves approximately 59.5\% in absolute execution time, all without compromising the optimization integrity.

\subsection{Calibration Accuracy Analysis}
\begin{figure}[htbp]
\centerline{\includegraphics[width=\columnwidth]{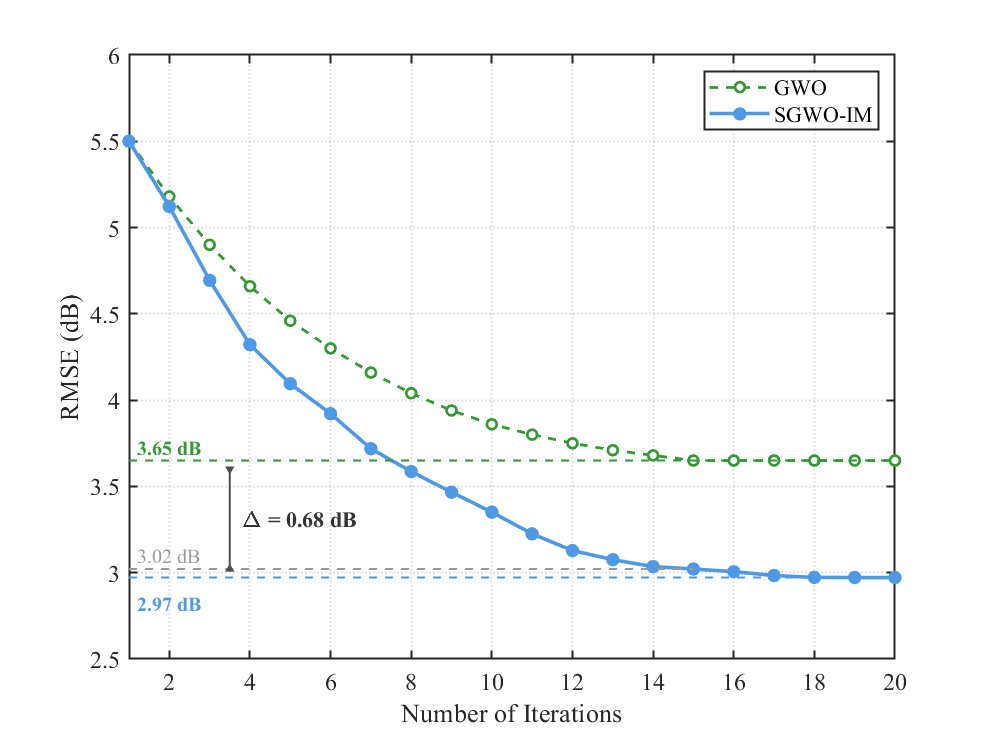}}
\caption{Comparison of convergence curves among different calibration algorithms.}
\label{fig_rmse}
\end{figure}

In typical urban and indoor RT scenarios, not all materials uniformly affect signal propagation. To ensure calibration efficiency and avoid high-dimensional overfitting, only the relative permittivities of three dominant materials, namely concrete, glass, and soil, were selected for optimization. These materials constitute the primary interacting surfaces, such as building walls, windows, and the ground, in the measurement environment, dominating the macroscopic reflection and scattering rays. The initial relative permittivities before calibration are strictly assigned based on the empirical material models recommended by the latest International Telecommunication Union (ITU) standard \cite{b21}, which serves as a widely adopted baseline in RT simulations. Table \ref{tab_materials} summarizes the relative permittivities optimized by the different algorithms. To demonstrate the algorithmic superiority in finding the global optimum, Fig. \ref{fig_rmse} visualizes the RMSE convergence behaviors of the two algorithms.

\begin{table}[htbp]
\caption{Relative Permittivities Before and After Calibration}
\begin{center}
\begin{tabular}{cccc}
\toprule
\textbf{Material} & \textbf{Before Calibration} & \textbf{GWO} & \textbf{SGWO-IM} \\
\midrule
Concrete & 5.31 & 6.45 & 5.82 \\
Glass & 6.27 & 4.12 & 4.65 \\
Soil & 2.25 & 3.88 & 4.21 \\
\bottomrule
\end{tabular}
\label{tab_materials}
\end{center}
\end{table}

As visually emphasized by the horizontal projection lines in Fig. \ref{fig_rmse}, the GWO suffers from premature convergence, completely stagnating at a local optimum of 3.65 dB after the 15th iteration, which is indicated by the green dashed line. Conversely, benefiting from the non-linear convergence factor and the individual history mechanism, SGWO-IM successfully escapes local traps. At the 15th iteration, SGWO-IM has already achieved a superior accuracy of 3.02 dB, represented by the gray dashed line, and continues to dig deeper into the EM parameter space, ultimately converging to 2.97 dB, marked by the blue dashed line. The explicit quantitative annotation highlights a significant final performance gap of 0.68 dB, proving the enhanced global search capability of SGWO-IM. Finally, the calibrated parameters detailed in Table \ref{tab_materials} were fed back into the RT engine to generate the simulated PL, which was then compared against the empirical measurement data. 

\begin{figure}[htbp]
\centering
\includegraphics[width=\columnwidth]{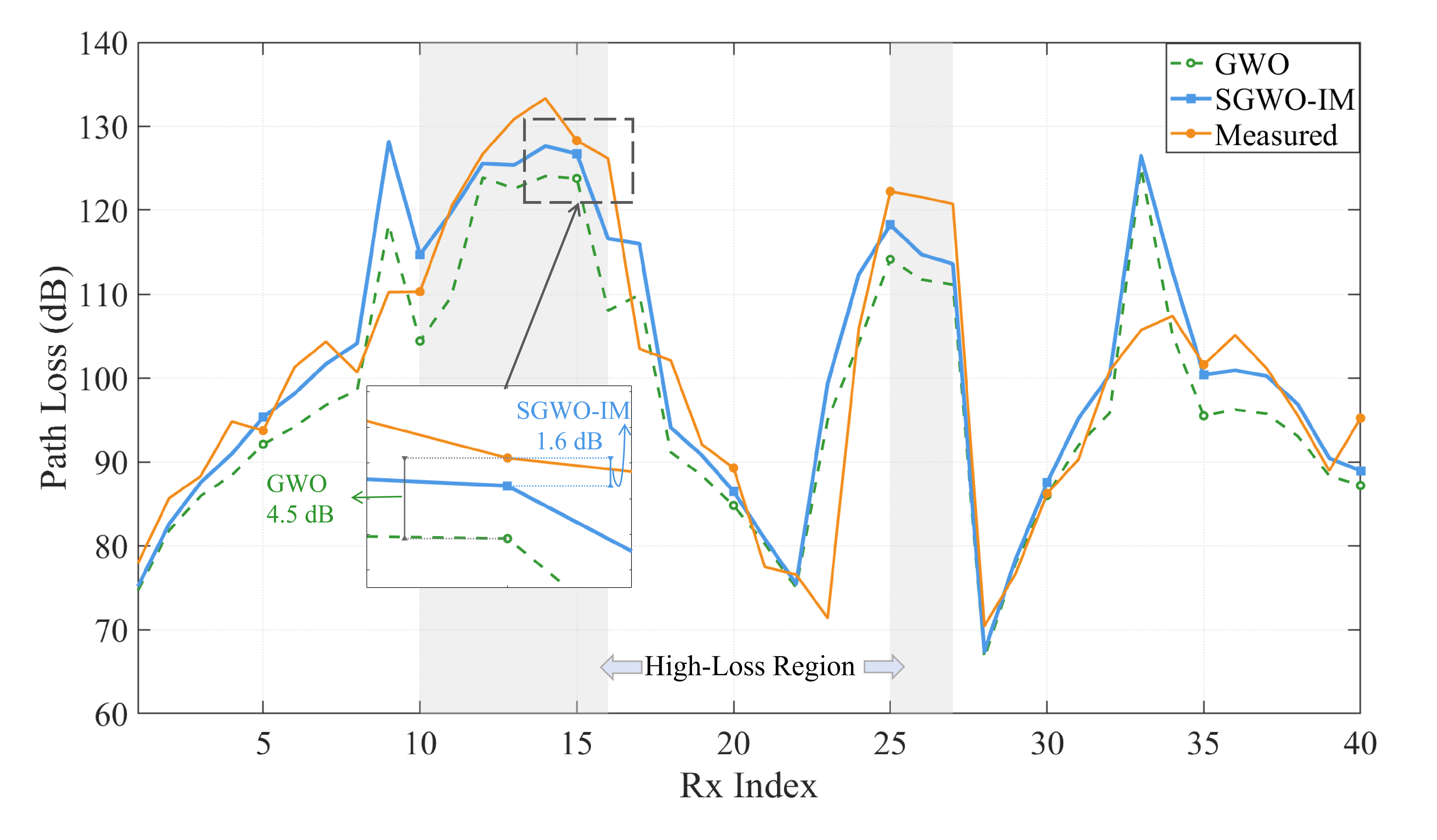}
\caption{Comparison between the simulated PL of different models and actual measurement data.}
\label{fig_pl}
\end{figure}

As depicted in Fig. \ref{fig_pl}, the SGWO-IM calibrated model exhibits exceptional consistency with the measured curve across all 40 Rx indices. This superiority is particularly evident in the High-Loss Regions, spanning from Rx 10 to 16 and Rx 25 to 27. As clearly highlighted by the zoom-in inset, in the High-Loss Region at Rx 15, GWO fails to capture the signal fluctuation, resulting in a significant local deviation of 4.5 dB. In contrast, SGWO-IM demonstrates superior robustness, firmly adhering to the measurement trend and suppressing this local deviation to 1.6 dB.

It is worth noting that at the specific minor point of Rx 9, the GWO appears to align slightly better with the measurement than SGWO-IM. This negligible anomaly occurs because the premature convergence of GWO coincidentally traps it in a localized parameter subset that matches the specific multipath superposition at this single location. However, SGWO-IM prioritizes the global spatial fidelity of the entire propagation environment, successfully preventing the model from overfitting to isolated noise points.

\section{Conclusion}
In this paper, we proposed the SGWO-IM algorithm for EM parameter calibration to support 6G digital twin channel reconstruction in complex urban environments. By integrating an online surrogate pre-screening mechanism with an individual memory strategy, this approach effectively overcomes the traditional bottleneck between computational efficiency and global calibration accuracy. Results indicate that the proposed method significantly accelerates the optimization process; by restricting the expensive RT evaluations to elite candidates, it reduces the required forward simulations by over 60\% compared to the standard GWO. Furthermore, the dynamic search mechanism successfully prevents premature convergence, allowing the algorithm to escape local optima traps. Consequently, the calibration framework achieves sub-3 dB precision, demonstrating a substantial improvement in global search capability. The calibrated EM parameters for dominant materials (concrete, glass, and soil) yield a simulated path loss that highly aligns with empirical measurements, exhibiting exceptional robustness particularly in deep-fading and High-Loss Regions. Ultimately, this research provides a highly efficient, practical, and accurate parameter calibration solution for 6G digital twin channel modeling and complex urban EM environment reconstruction.

\section*{Acknowledgment}
This work was supported in part by National Natural Science Foundation of China (62571053, 62525101, 62341128), National Key Research and Development Program of China (2023YFB2904805), Beijing Municipal Natural Fund (L243002), and BUPT-CMCC Joint Institute.

\end{document}